\documentclass [a4paper,fleqn] {cas-sc}

\usepackage[numbers]{natbib}
\def\tsc#1{\csdef{#1}{\textsc{\lowercase{#1}}\xspace}}
\tsc{WGM}
\tsc{QE}
\usepackage[utf8]{inputenc}
\usepackage{float}
\usepackage{placeins}

\begin{document}
\let\WriteBookmarks\relax
\def\floatpagepagefraction{1}
\def\textpagefraction{.001}

\newcommand\blfootnote[1]{%
  \begingroup
  \renewcommand\thefootnote{}\footnote{#1}%
  \addtocounter{footnote}{-1}%
  \endgroup
}

\shorttitle{\textcolor{red}{This is the author's version which has not been fully edited and contents may change prior to final publication. Citation information: 10.1016/j.compbiomed.2022.105686}}

\shortauthors{\textit{This work is licensed under a Creative Commons Attribution-NonCommercial-NoDerivatives (CC-BY-NC-ND) 4.0 License. For more information, see https://creativecommons.org/licenses/by-nc-nd/4.0/}}
\title [mode = title]{A Nonlinear Beamforming for Enhanced Spatiotemporal Sensitivity in High Frame Rate Ultrasound Flow Imaging}

\author[1]{A. N. Madhavanunni}

\cortext[1]{\textcolor{red}{Article accepted for publication in Elsevier Computers in Biology and Medicine. Citation information: 10.1016/j.compbiomed.2022.105686}\\\textcolor{red}{This is the author's version which has not been fully edited and contents may change prior to final publication.}}





\affiliation[1]{organization={Center for Computational Imaging, Department of Electrical Engineering},
            addressline={Indian Institute of Technology (IIT) Palakkad}, 
            country={India}}

\author[1]{Mahesh Raveendranatha Panicker}







\begin{abstract}
Typically, an ultrasound flow imaging system employs the conventional delay and sum (DAS) beamformer due to its inherent low complexity. But the conventional DAS technique offers poor contrast, low imaging resolution, and limited spatiotemporal sensitivity.This article attempts to improve the spatiotemporal sensitivity of the conventional flow imaging with a novel multiply and sum based nonlinear high resolution (NLHR) beamforming approach. The major advantages of the proposed beamformer are the harmonic generation and the enhanced coherence in beamformed signals that improve the spatiotemporal sensitivity towards flow transients. We demonstrate the proposed beamformer for a directional cross-correlation as well as an autocorrelation based velocity estimator with simulated parabolic flow profiles of different velocities and flow directions, an \textit{in-vitro} rotating disk dataset, and pulsatile flow experiments. The sensitivity of NLHR beamforming towards the flow transients is validated \textit{in-vitro} with a sudden reversal of flow direction and air bubble tracking experiments.The comparison between the time-frequency plots of DAS and NLHR beamforming indicates that the impulsive spatiotemporal changes induced by the flow of air bubbles were clearly characterized by nonlinear beamforming than that of DAS beamforming. Furthermore, better spatiotemporal velocity tracking of a single air bubble and a clear distinguishability between the tracking of two proximal air bubbles were observed \textit{in-vitro}. Preliminary studies on the \textit{in-vivo} carotid data also show comparable, if not better, results than that of the DAS algorithm. Detailed results for each test case in simulation, phantom, and \textit{in-vivo} studies are available as movies with the supplementary material and online [\href{https://www.youtube.com/playlist?list=PLiuuVhVNWBZSYikqhd20FsVr8NTKRlZ4F}{Online Link}]. 

\end{abstract}


\maketitle

\section{Introduction}\label{Introduction}
Ultrasound Doppler technique is the most widely used non-invasive technique for flow imaging and has a great diagnostic value in investigating vascular hemodynamics  \cite{Nelson1988TheMean}. It uses high frequency non-ionizing sound waves to create multiple insonification of blood vessels which makes it the safest technique to quantify the complex flow dynamics. Traditional Doppler ultrasound-based flow imaging techniques are highly dependent on the beam-to-flow angle and least sensitive to the transverse flow component \cite{Nelson1988TheMean, Tortoli1993TransverseSonograms,Newhouse1994Three-DimensionalWidth}. This makes it difficult to quantify the complex flows associated with the superficial vasculature as they are immediately beneath the skin surface and the flow is purely transverse in most cases.

To address these limitations, a multi-dimensional estimation for velocity vectors, commonly known as vector flow imaging (VFI) has been proposed \cite{Jensen1998AVectors}. It provides an angle-independent, absolute quantification of the flow rate and enables the dynamic multi-dimensional visualization of complex blood flow. In this regard, various methods like transverse oscillation \cite{Jensen1998AVectors}, crossbeam Doppler \cite{Fox1978MultipleVelocimetry}, speckle tracking \cite{GreggE.Trahey1987AngleFlow}, vector triangulation \cite{Overbeck1992VectorDimensions}, directional beamforming \cite{Jensen2003DirectionalSimulation, Jensen2003DirectionalInvestigation}, and its variants coupled with autocorrelation and cross-correlation based velocity estimation techniques have been exploited \cite{Jensen2019EstimationTheory,Jensen2019EstimationInvestigation,Fadnes2015RobustImaging,Ricci2014Real-timeWaves,Yiu2014VectorPatterns,Yiu2016Least-SquaresImaging,Lenge2015Plane-waveImaging,Ricci2018Real-TimeRegion,Madhavanunni2020TriangulationFlows}. Most of the above strategies make use of unfocused transmit schemes like diverging waves \cite{Jensen2019EstimationTheory,Jensen2019EstimationInvestigation}, multi-angle steered plane waves \cite{Fadnes2015RobustImaging,Ricci2014Real-timeWaves,Yiu2014VectorPatterns,Yiu2016Least-SquaresImaging}, or single non-steered plane waves \cite{Lenge2015Plane-waveImaging,Ricci2018Real-TimeRegion,Madhavanunni2020TriangulationFlows,madhavanunni2021angle} to insonify the region of interest at high frame rates. 

The directional beamforming technique in \cite{Jensen2003DirectionalSimulation} and \cite{Jensen2003DirectionalInvestigation} focuses the received signals along the flow direction and employs cross-correlation to estimate the velocity magnitude. But this approach requires the knowledge of the flow angle with respect to the emitted beam for beamforming. In this regard, correlation and numerical triangulation-based techniques have been developed to estimate the flow angle \cite{Kortbek2006EstimationMethod,Hoyos2016AccurateImaging}. However, synthesizing the directional signals by rotating the grid would be difficult in the case of complex flows in which the flow is not in a well-defined direction. To address these limitations, vector triangulation-based flow imaging approaches are proposed in the literature \cite{Yiu2014VectorPatterns,Yiu2016Least-SquaresImaging,Lenge2015Plane-waveImaging,Ricci2018Real-TimeRegion,Madhavanunni2020TriangulationFlows}. However, the contrast, resolution and accuracy of the velocity estimation are mainly determined by the beamforming algorithm employed in the system.

Delay and sum (DAS) is the most commonly used beamforming technique in medical ultrasound imaging due to its inherent low complexity \cite{Perrot2021SoBeamforming,Parker2013Correspondence:Functions}. Conventional DAS beamformer in ultrasound systems involves delaying the received signals and then summing it up to obtain the dynamically focused receive beams. Usually, the beamformer uses an apodization window to weight the delay compensated signal before summing. The apodization enhances the energy of the signal components from the desired grid point while suppressing the signal components from the undesired locations \cite{BeenLim2008ConfocalAlgorithm}. This effectively improves the signal-to-noise ratio and provides a better beam at the output of the beamformer \cite{Parker2013Correspondence:Functions}. But the conventional DAS technique employed in most of the aforementioned flow imaging methods offers poor contrast, low imaging resolution, and limited spatiotemporal sensitivity. In this paper, the term sensitivity is intended as the capability to detect the flow velocity changes and the transient events.  

Improving the spatial resolving capability of VFI helps in understanding more localized flow patterns and velocity distribution in various pathological conditions where VFI is used for diagnosis and monitoring. Enhanced sensitivity in VFI towards highly dynamic and impulse components in velocity aids in identifying new biomarkers or pathological features in medical diagnostics. Moreover, the importance of enhancing the sensitivity towards impulse velocity components and higher temporal resolution (reduced time window to detect velocity changes) is further emphasized in various applications such as functional ultrasound imaging of the brain \cite{Dizeux2019FunctionalPrimates}, microbubble localization and tracking, remote drug delivery, neovascularization imaging \cite{Tanter2014UltrafastUltrasound,Yu2018Super-resolutionAccuracy}. 
\subsection{Related Work}
One of the early attempts towards improving the sensitivity of conventional Doppler has been the coherent plane wave compounding (CPWC) at high-frame rates as reported in \cite{Tanter2014UltrafastUltrasound,bercoff2011ultrafast}. The CPWC technique employs multiple steered plane wave transmit and coherently combines the beamformed data to obtain a high-resolution image and to estimate the velocity \cite{bercoff2011ultrafast,jensen2016ultrasound}. However, this technique was predominantly employed for power Doppler imaging to increase the sensitivity of axial velocity components. The influence of motion in CPWC has been studied by Denarie, Bastien et al., and a compensation scheme for radial motion is reported in \cite{denarie2013coherent}, but the effect of multi-directional velocity components is not addressed.  Moreover, with the increase in the number of steered plane wave transmit, the effective frame rate reduces and hence the CPWC based approach would be ideal for the investigation of low velocity flows as shown in \cite{mace2011functional} and \cite{mace2013functional}. 

Later, singular value decomposition (SVD) based filters were introduced to increase the spatiotemporal sensitivity in high frame rate Doppler imaging \cite{Demene2015SpatiotemporalSensitivity,baranger2018adaptive}. Subsequently, principal component analysis based approaches have been demonstrated \textit{in-vivo} for vascularization imaging \cite{shen2019high} and surgical management of open-brain tumors \cite{barthelemy2019development}. Recently, machine learning and deep learning based approaches are emerging as automated ultrasound image analysis tools and have shown promising results for various diagnostic imaging tasks like detection, segmentation, and classification of tumors and lesions \cite{liu2019deep, gharaibeh2022radiology, shehab2022machine}. The convolutional neural networks, despite being successful in optical flow estimations, have not been exploited much for ultrasound blood flow imaging and very limited efforts have been reported in the literature \cite{stanziola2018deep,li2019vector,park2022ultrasound}. Moreover, most of the efforts in flow imaging have been coupled with ultrasound contrast agents and are focused on obtaining the vascular maps at a high spatial resolution but don’t contribute much towards improving the temporal sensitivity or the temporal super-resolution \cite{errico2015ultrafast,cohen2016ultrasensitive,chen2022deep}. However, it has been reported that the deep learning models could provide better results by incorporating temporal information of the microbubbles \cite{chen2022deep}.

Towards improving the temporal resolution, a minimum variance based adaptive velocity estimation technique has been recently reported in \cite{Makouei2020AdaptiveStudy}, but with higher computational complexity. However, the efforts towards employing nonlinear beamformers of lower computational complexity (than minimum variance) like delay multiply and sum (DMAS) technique for flow imaging is limited. Delay multiply and sum, commonly known as DMAS algorithm is a nonlinear beamforming technique in B-mode ultrasound imaging which has a higher spatial and contrast resolution with a better clutter and noise rejection when compared to DAS \cite{BeenLim2008ConfocalAlgorithm,Matrone2015TheImaging}. In one of the recent studies reported in \cite{Matrone2015TheImaging} the DMAS algorithm is modified to a filtered delay multiply and sum (F-DMAS) by employing an additional bandpass filter for Brightness-mode (B-mode) imaging. The simulation and the \textit{in-vivo} studies with F-DMAS algorithm have demonstrated a better-quality image with a higher dynamic range and improved contrast resolution at the output \cite{Matrone2017HighBeamforming}. This nonlinear technique is further investigated for plane wave and synthetic aperture high frame rate ultrasound imaging \cite{Matrone2017HighBeamforming,Mozaffarzadeh2018Double-StageImaging} as well as for photoacoustic imaging \cite{Mozaffarzadeh2018Double-stageImagingb,Park2016Delay-multiply-and-sum-basedMicroscopy}. Similar to the formulation of DMAS, spatial coherence based techniques have been proposed for better clutter suppression in power Doppler imaging \cite{Dahl2013CoherentCoherence,Li2015CoherentBeamforming,ozgun2019spatial}. In this regard, the quality improvement in B-mode and power Doppler imaging with DMAS based beamforming \cite{Matrone2017HighBeamforming} could be leveraged to improve the resolution and sensitivity of VFI techniques.

\subsection{Objective and Contributions}
Inspired by the F-DMAS beamforming in B-mode imaging, this paper attempts to address the spatiotemporal sensitivity of the conventional flow imaging techniques with a novel non-linear beamforming approach without the use of any contrast agents and deep learning based methods. The major contributions of this work are as follows. 
\begin{enumerate}
    \item A novel nonlinear beamforming technique is proposed for ultrasound flow imaging and to the best of the authors’ knowledge, this is the first effort towards the application of nonlinear beamforming in flow imaging.
    \item We validate the proposed nonlinear beamformer using typical parabolic flow simulations with a cross-correlation based velocity estimator and an autocorrelation based velocity estimator which are the common velocity estimation techniques in the literature. 
    \item Further, the proposed approach has been thoroughly investigated for velocity sensitivity with \textit{in-vitro} datasets including a rotating disk, air bubble tracking, and flow direction reversal. Finally, we report the \textit{in-vivo} performance evaluation of the proposed approach for a typical pulsatile flow in a carotid artery dataset.
    \item The results, when compared to the state-of-the-art DAS based flow imaging approaches, suggest that the proposed method provides better spatiotemporal sensitivity towards the flow transients.
\end{enumerate}

The rest of the article is organized as follows. The following section (Section \ref{proposed_approach}) explains the proposed nonlinear high-resolution beamforming technique for flow imaging and describes how it can be used in conjunction with triangulation with auto-correlation estimator and directional cross correlation estimator. Section \ref{implementation} discusses the implementation details along with the simulation and experimental setup. Section \ref{results} presents the results obtained from the simulations and experimental investigations in comparison with DAS based VFI methods and the discussion of results is presented in Section \ref{discussion}. The article is concluded in Section \ref{conclusion} with a summary of the work.

\section{Proposed Approach}\label{proposed_approach}
A schematic of the proposed non-linear high resolution (NLHR) beamformer for flow imaging is shown in Fig. \ref{fig1}(a). A detailed description of each step in the approach is provided in the subsequent sections.
\subsection{Nonlinear High-Resolution Beamformer}\label{NLHRBF}
The first part of the NLHR beamforming deals with dynamic delay compensation of the received signals to enable channel-wise spatial focusing. The second part consists of the design of an appropriate apodization function to form the directional weighted focused beams at left and right receive sub-apertures. Further, these beams are multiplied pairwise and summed before it is filtered to obtain the higher harmonic beamformed signals in the final step. These beams are used to estimate the velocity magnitude and direction. It must be noted that these beams will be having higher lateral resolution and better contrast (clutter suppression) as discussed in \cite{Matrone2015TheImaging}. In this work, non-steered plane waves are used to insonify the region of interest at very high frame rates to aid the improvement in velocity sensitivity. However, the proposed approach is also applicable to multi-angle plane waves, synthetic aperture and diverging wave transmit schemes. A detailed explanation of the approach is provided in the following subsections.
\subsubsection{Channel Directive Beam Synthesis}\label{CDBS}
\begin{figure*}[pos=!t]
	\centering
	  \includegraphics[width=\textwidth,height=\textwidth,keepaspectratio]{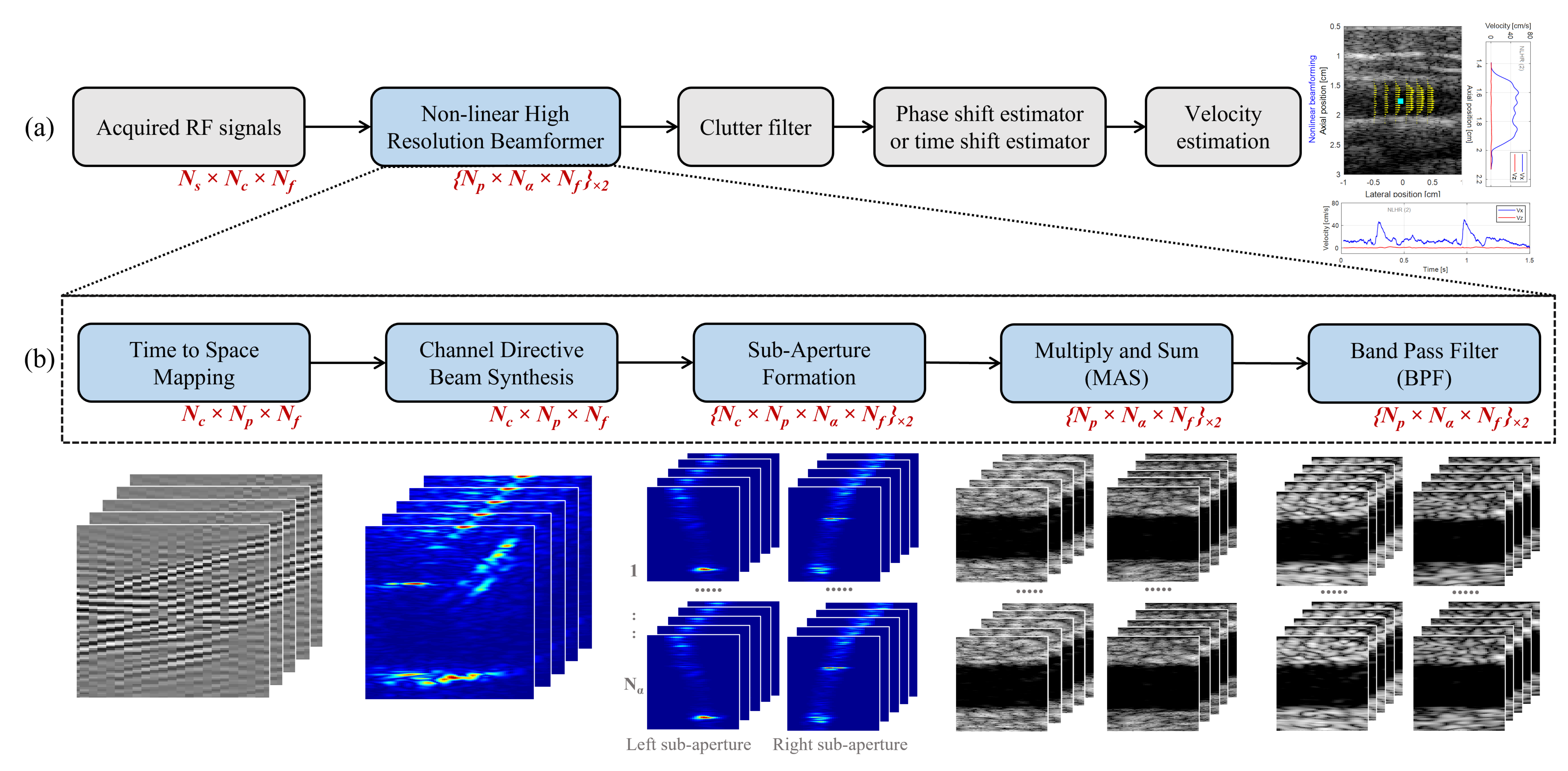}
	\caption{(a) Illustration of the proposed non-linear high-resolution (NLHR) beamformer for flow imaging application with typical velocity estimation schemes. (b) Block diagram illustration of the proposed nonlinear beamformer. The size of the matrix after each operation is shown at the bottom of each block. x2 indicates 2 matrices, one for left sub-apertures and other for right sub-apertures.}
	\label{fig1}
\end{figure*}

The received radio frequency (RF) echo signals, $e_i(t)$ (where $i$ represents the $i^{th}$ transducer), are delay compensated and mapped to a two-dimensional grid in the region of interest. This process is referred to as time to space mapping of the RF signals. The delay compensated signal, $e_i(p)$ for a given pixel p is obtained as $e_i\left(\tau_{i,p}\right)$, where, $\tau_{i,p}$ is given by \eqref{eqn0}:
\begin{equation}\label{eqn0}
\tau_{i,p} = \frac{1}{c}\left\{z_p+\sqrt{{(x_i-x_p)}^2+{(z_i-z_p)}^2}\right\}
\end{equation}
where $\left\{x_i,z_i\right\}$, $\left\{x_p,z_p\right\}$ are the transducer and the pixel coordinates (corresponding to pixel $p$) respectively. It is assumed that a linear array transducer is placed at the start depth (i.e., at $z_i=0$) and $c$ is the speed of sound. The size of the delay compensated signals for the full region of interest will be $N_c \times N_p \times N_f$, where, $N_c$ is the number of channels, $N_p$ is the number of pixels and $N_f$ is the number of frames. The non-steered plane wave excitation is assumed. The channel directive beams for a pixel, $s_i(p)$, are formed with each of the transducers as the beam centers according to \eqref{eqn1}. 
\begin{equation}\label{eqn1}
    s_i(p)=\ \sum_{j=1}^{N_c}{{W}_{i,p}(j)}\ e_i\left(\tau_{i,p}\right)
\end{equation}
where, $s_i\left(p\right)$ is the apodized amplitude of the $i^{th}$ transducer for $p^{th}$ pixel, ${W}_{i,p}$ is a Gaussian weight vector, centered at $i^{th}$ receive element. The width of the Gaussian function is dynamically sized according to the receive distance (i.e., $width\ =\ receive\ distance/F_n$, where $F_n$ is the F-number), instead of depth as in traditional apodization. Thus, unlike traditional apodization, the channel directive beam synthesis as given by \eqref{eqn1} creates $N_c$ beam observations, that are equivalent to the focused observations from $N_c$ different angles for each pixel. The size of the matrix $S$ would be $N_c \times N_p \times N_f$ as shown in Fig. \ref{fig1} and obtained by repeating \eqref{eqn1} for $N_p$ pixels and $N_f$ frames. This step converts the delay compensated channel data to the spatially focused channel beams and its significance in flow imaging has been evaluated in our previous work \cite{Madhavanunni2020DirectionalImaging}.

\subsubsection{Sub-Apertures Formation}\label{SAF}
The next step is to apply suitable weight onto the channel directive beams to obtain multiple sub-apertures to the left and right of the selected transmit element following the principles of triangulation. The formation of sub-apertures is a traditional step involved in any the triangulation-based velocity estimation scheme \cite{Overbeck1992VectorDimensions,Ricci2014Real-timeWaves,Yiu2014VectorPatterns,Madhavanunni2020TriangulationFlows}. These are referred to as left and right sub-aperture signals, ${A_{L_{\alpha,i}}}(p)$ and ${A_{R_{\alpha,i}}}(p)$, corresponding to different transmit-receive angles ($\alpha$) and are given by \eqref{eqn2} and \eqref{eqn3}:
\begingroup
\begin{equation}\label{eqn2}
    {A_{L_{\alpha,i}}}(p)=\ {{W}_L}_{\alpha,p}(i)s_i(p)	
\end{equation}
\begin{equation}\label{eqn3}
    {A_{R_{\alpha,i}}}(p)=\ {{W}_R}_{\alpha,p}(i)s_i(p)
\end{equation}
\endgroup
where, ${{W}_L}_{\alpha,p}$ and ${{W}_R}_{\alpha,p}$ are the left and right sub-aperture apodization weight vector of length $N_c$. For each pixel, a Gaussian weight vector centered at the receive element that corresponds to the transmit-receive angle ($\alpha$) under consideration is used. This provides maximum weightage to the beam received at an angle $\alpha$ and thus ensures the directionality of the left and right receive sub-aperture signals corresponding to each pixel in the region of interest. Unlike in the traditional DAS beamformer, the weighted signals are not directly summed in our approach and hence the signals obtained with \eqref{eqn2} and \eqref{eqn3} are regarded as $\alpha$-weighted beam focused left and right sub-aperture signals. It should be noted that these signals are weighted in the direction defined by the $\alpha$ and not the flow direction. This sub-aperture formation at receive enables the beamforming to be employed for most of the split aperture vector flow imaging techniques. 
\subsubsection{Multiply and Sum Operation}\label{MAS}
Once the $\alpha$-weighted beam focused left and right sub-aperture signals are formed, each beam is multiplied with every other beam and then summed up to obtain the equivalent DMAS beamformed signal in VFI, $\hat{y}_{L_\alpha}\left(p\right)$ and $\hat{y}_{R_\alpha}\left(p\right)$, as given by \eqref{eqn4} and \eqref{eqn5}.
\begingroup
\begin{equation}\label{eqn4}
    {{\hat{y}}_{L_\alpha}}\left(p\right)=\ \sum_{i=1}^{N_c-1}\sum_{j=i+1}^{N_c}{{A_L}_{\alpha,i}(p)\times{A_L}_{\alpha,j}(p)}
\end{equation}
\begin{equation}\label{eqn5}
    {{\hat{y}}_{R_\alpha}}\left(p\right)=\ \sum_{i=1}^{N_c-1}\sum_{j=i+1}^{N_c}{{A_R}_{\alpha,i}(p)\times{A_R}_{\alpha,j}(p)}
\end{equation}
\endgroup
It can be observed that the multiply and sum operation is mathematically a cross-correlation function between all the receive channels in which the self-product terms $(i = j)$ are excluded. Also, it is to be noted that either $[{A_L}_{\alpha,i}(p)\times{A_L}_{\alpha,j}(p)]$ or $[{A_L}_{\alpha,j}(p)\times{A_L}_{\alpha,i}(p)]$ is considered in the summation of multiply and sum operation as both the coefficients are equal in magnitude and hence the information is redundant. This will halve the corresponding correlation coefficient and reduce the number of required multiplications by a factor of 2 \cite{Matrone2015TheImaging,Ramalli2017HighBeamforming}. The total number of multiplications required for an N channel receive aperture, $M_N$, is given by \eqref{eqn6}.
\begin{equation}\label{eqn6}
    M_N=\ \frac{N^2-\ N}{2}
\end{equation}
\subsubsection{Band Pass Filtering}
The frequency spectrum of the delay compensated directional focused beams of each channel will be a band centered around the center frequency of the insonified plane wave, $f_0$. But the frequency spectrum of multiply and sum beamformed signals, ${\hat{y}_{L_\alpha}}\left(p\right)$ and ${\hat{y}_{R_\alpha}}\left(p\right)$, will have two peaks, one centered at $f_0-\ f_0\ \approx\ 0$ and the other at $f_0+\ f_0\ \approx\ 2f_0$ because of the fact that \eqref{eqn4} and \eqref{eqn5} multiply two signals having almost similar frequency \cite{Matrone2015TheImaging}. Now, a band pass filter with a passband centered around $2f_0$ is used in the final step to remove the DC component. Thus, the frequency spectrum of filtered left and right beamformed signals, ${{\widetilde{y}}_{L_\alpha}}\left(p\right)$ and ${{\widetilde{y}}_{R_\alpha}}\left(p\right)$, will be centered at $2f_0$. In any Doppler technique, the higher the transmit frequency, the higher the Doppler shift and the better the velocity estimate. In this work, since the harmonics induced by multiply and sum is employed, it is expected to give a higher Doppler resolution.
\subsection{Velocity Estimation}
A truncated SVD based spatiotemporal clutter rejection filter \cite{Demene2015SpatiotemporalSensitivity} (as available in the Matlab UltraSound Toolbox \cite{Posada2016StaggeredDoppler}) is used to remove the wall motion. The details of the SVD filter are provided in the results section. As shown in Fig. \ref{fig1}(a), the flow velocity vectors can be derived either by employing an autocorrelation-based estimator or a cross-correlation based estimator. Both the methods are described briefly in the following subsections.

\subsubsection{Triangulation with Autocorrelation (TAC)}\label{TAC}
The mean slow-time frequency of the clutter suppressed left and right beamformed signals, $y_{L_\alpha}\left(p\right)$ and $y_{R_\alpha}\left(p\right)$, is determined using the classical lag-one autocorrelation approach \cite{Kasai1985Real-TimeTechnique}. A window size of $k$ is used to estimate the autocorrelation. Once the slow-time frequency, $f_{L_\alpha}$ and $f_{R_\alpha}$, are estimated, the orthogonal components of the velocity can be estimated using the following equations \eqref{eqn7} and \eqref{eqn8} as in conventional triangulation \eqref{eqn8} \cite{Overbeck1992VectorDimensions} and that used in \cite{Ricci2018Real-TimeRegion}:
\begin{equation}\label{eqn7}
    v\ cos\ \theta\ =\frac{f_{L_\alpha}+f_{R_\alpha}}{(1+cos\ \ \alpha)}\times\frac{c}{2f^\prime}
\end{equation}
\begin{equation}\label{eqn8}
    v\ sin\ \theta\ =\frac{f_{L_\alpha}-{f_{R_\alpha}}}{(sin\ \ \alpha)}\times\frac{c}{2f^\prime}
\end{equation}
where $f^\prime = f_0$ in the case of DAS based beamforming and $f^\prime= {2f}_0$ in the case of DMAS based beamforming because the center frequency of the DMAS based beamformed signal is around ${2f}_0$. Unlike in \cite{Ricci2018Real-TimeRegion} where only one pair of observations is considered for triangulation, the multiple sub-aperture formation in the proposed approach provides multiple observations enabled by multiple $\alpha$ at receive. The velocity magnitude and flow direction are calculated as the mean of all such observations.

\subsubsection{Directional Cross-Correlation (DCC)}\label{DCC}
Directional cross-correlation based velocity estimation as in \cite{Jensen2003DirectionalSimulation} is performed for the observations enabled by multiple $\alpha$ values at receive and the mean estimates among them are computed. However, to perform DCC, beamforming must be done along the flow direction. This is done by rotating the flow grid and the time to space mapping of the received RF signals is done for the rotated spatial grid. Apart from this, no changes are required in the beamformer architecture to perform DCC based velocity estimation. A correlation interval of $-10\lambda:10\lambda$ (equivalently, a cross-correlation window size of $L=20\lambda$) and a sampling interval of $0.1\lambda$ are chosen for DCC, as in \cite{Jensen2003DirectionalSimulation}.

\section{Implementation and Validation}\label{implementation}
In this section, the details on the implementation of the proposed approach in MATLAB® as well as the simulation and experimental setup used for the evaluation and validation of the proposed approach are discussed.

\subsection{Considerations to Avoid Aliasing}
With the multiply and sum operation discussed in Section \ref{MAS}, the frequency spectrum of the beamformed signal would get shifted to $2f_0$ and the Doppler shift would also get doubled. To ensure no aliasing due to the frequency doubling, the received RF data is resampled along the axial direction as well in the temporal (frame) direction immediately after the acquisition to satisfy the Shannon-Nyquist sampling theorem. The size $N_c \times N_s \times N_f$ in Fig. \ref{fig1} represents the size of the received RF signals after resampling. It is assumed that the acquired RF data is not aliased and has a sufficient frame rate to capture the flow velocities as in any Doppler system acquisition. 


\begin{table*}[width=.8\textwidth,cols=4,pos=t]
\caption{Parameters used for simulation and experimental studies}\label{tbl1}
\begin{tabular*}{\tblwidth}{LCLL}
\hline \hline
\multicolumn{1}{L}{\multirow{2}{*}{Parameter}} & \multicolumn{1}{c}{\multirow{2}{*}{Symbol}} & \multicolumn{2}{c}{Value}\\ \cline{3-4} 
\multicolumn{1}{c}{}& \multicolumn{1}{c}{}& \multicolumn{1}{L}{Simulation} &\multicolumn{1}{L}{Experiment} \\ \hline
\multicolumn{4}{L}{Probe and transmit parameters}\\ \hline
Type &  & Linear    & Linear (L11-5v)\\ 
No. of elements& $N$ & 128   & 128   \\
Element pitch  & $p$ & 0.1925 mm & 0.3 mm  \\
Center frequency   &  $f_0$  & 8 MHz & 7.6 MHz \\
Pulse repetition frequency  & $f_{PRF}$   & 10 kHz & 8 kHz \\
Mode     &       & Plane wave & Plane wave\\
No. of transmit cycles   &  & 5  & 5 \\
Apodization  &     & Rectangular & Rectangular \\ \hline
\multicolumn{4}{L}{Phantom parameters}                                        \\ \hline
Speed of sound &  $c$  & 1540 m/s     & 1540 m/s  \\
Wavelength  & $\lambda$  & 0.1925 mm & 0.2026 mm  \\
Inner diameter of vessel   & $d_i$  & 10 mm  & 4.7625 mm  \\
Outer diameter of vessel  &  $d_0$  & 10 mm & 6.35 mm  \\
Peak velocities of flow   &   $v_0$   & 5 - 100 cm/s & 5 - 50 cm/s \\
Scatterer density &   $D$     & 10 per $\lambda^3$ &   \\ \hline
\multicolumn{4}{L}{Receive Beamforming Parameters}                              \\ \hline
Sampling frequency   &   $f_s$   & 100 MHz  & 31.25 MHz \\
Transmit - Receive angles & $\alpha$ & $6^\circ, 9^\circ, 12^\circ, 15^\circ$ & $6^\circ, 9^\circ, 12^\circ, 15^\circ$ \\
F-number   & $F_n$ & 1.25   & 1.7   \\
Apodization  &  & Gaussian  & Gaussian  \\  
\hline \hline
\end{tabular*}
\end{table*}

\subsection{Simulation Setup}
The performance of the proposed approach is initially evaluated and validated with extensive simulations using Field II \cite{Jensen1992CalculationTransducers,jensen1996field} and MATLAB R2019a (The MathWorks Inc., Natick, MA, USA). Flow phantoms are generated using the Field II program (version 3.24) with the parameters shown in Table \ref{tbl1}. A standard linear array probe with 128 active elements is modeled in the Field II simulator and no transmit apodization is applied for simulation studies. A 5-cycle sinusoid at $8\ MHz$ is used for transducer excitation and a Hanning weighted single cycle sinusoid is used as the transducer impulse response at the transmit and receive. Non-steered plane wave transmit is used for all the datasets in this paper. The simulation parameters closely follow the parameters chosen in \cite{Jensen2003DirectionalSimulation}.

The performance of the NLHR beamforming with TAC and DCC based velocity estimators is compared to the true/theoretical profiles and the corresponding velocity estimators with DAS beamforming. The DAS beamforming for the TAC velocity estimator adopts the single transmit dual-angle multi receive scheme (which is proposed in one of our recent works \cite{Madhavanunni2020TriangulationFlows}). The DAS beamforming for the DCC velocity estimator adopts the directional beamforming scheme reported in \cite{Jensen2003DirectionalSimulation}. To have a fair comparison, the transmit-receive angles, correlation interval, and other parameters are kept the same for DAS and NLHR beamforming. The resampled RF data corresponding to $L=20\lambda$ and $k=0.8\ ms$ is used for DAS as well as NLHR beamforming unless otherwise mentioned. It should be noted that the comparison of NLHR and DAS beamforming for DCC is only done for magnitude estimates as DCC estimates only provide the velocity magnitude. 

Three different test cases were investigated with simulations for the performance evaluation.

\textit{Test case 1:} To verify and validate the accuracy in velocity estimation, typical parabolic flow profiles, as given by \eqref{eqn9} \cite{Jensen2003DirectionalSimulation}, having the peak velocities, $V_P = 5$, 10, 25, and $50\ cm/s$ is simulated for a transverse vessel of radius, $R = 5\ mm$ with its axis at $25\ mm$ depth from the probe surface. 
\begin{equation}\label{eqn9}
    v_p\left(r\right)=\ V_P\left[\ 1\ -\ \frac{r^2}{R^2}\right]	
\end{equation}
where r is the radial position in the vessel. The velocity magnitude and angles are estimated with TAC and DCC for NLHR beamforming and are compared with the corresponding estimates with DAS beamforming and the true profiles. 

\textit{Test case 2:} In the second test, datasets are simulated for a $5\ mm$ radius vessel whose axis is inclined at $-20^\circ$, $-10^\circ$, $0^\circ$, $10^\circ$, and $20^\circ$ with the horizontal and having a peak velocity, $V_P = 50\ cm/s$. The performance comparison is done as in test case 1. 

\textit{Test case 3:} The third test aims to study the influence of correlation intervals in TAC and DCC estimators with DAS and NLHR beamforming. The correlation interval for DCC for comparison is chosen as $-10\lambda:10\lambda$ and $-5\lambda:5\lambda$, denoted by $L=20\lambda$ and $10\lambda$, respectively whereas the averaging interval is chosen as $k=0.8\ ms$ and $1.6\ ms$. The correlation interval for TAC for comparison is chosen as $k=0.8\ ms$ and $1.6\ ms$ whereas the averaging interval is chosen as $L=20\lambda$ and $10\lambda$. The results for the 4 combinations of the $k$ and $L$ values are compared for DAS and NLHR beamforming with TAC and DCC estimators. 

\subsection{Experimental Setup}
\textit{In-vitro Rotating Disk}: The feasibility of the proposed approach is verified \textit{in-vitro} by using a rotating disk (diameter of $2\ cm$) dataset \cite{Madiena2018ColorSampling,Garcia2021MakeToolbox}. The dataset was acquired using a Verasonics scanner (Verasonics Inc., Kirkland, WA, USA) and a 128-element linear array with non-steered plane wave insonification at a center frequency of $5\ MHz$ and a pulse repetition frequency of $10\ kHz$. The RF signals for 32 frames with a sampling frequency of $20\ MHz$ were recovered from the bandpass sampled (at a sub-Nyquist sampling rate of $6.66\ MHz$) RF dataset before beamforming. 

\textit{In-vitro Flow Phantom Experiments}: The proposed approach is thoroughly investigated for velocity sensitivity using a commercial Doppler ultrasound flow simulator from CIRS (CIRS Inc., Norfolk, VA, USA). The simulator includes a peristaltic flow Pump (Model 769), Doppler fluid (Model 769DF), and a tissue-mimicking Doppler ultrasound flow phantom (Model 069A) which has a peripheral vessel with a controlled flow velocity. Three test cases with pulsatile flow profiles were investigated in the flow phantom at a pulse repetition frequency of $8\ kHz$. For the first test case, a typical pulsatile flow with a peak velocity of $50\ cm/s$ is configured. In the second test case, the Doppler fluid is seeded with some air bubbles by gently shaking the fluid and the pulsatile flow with parameters same as above is configured using the flow pump. The velocity sensitivity and spatial localization capability are evaluated by tracking air bubbles over time with the proposed NLHR approach. This aims to test the sensitivity of the proposed approach towards the impulsive changes in the flow induced by the air bubbles. In the third test case, a pulsatile flow with a sudden reversal of direction is imposed during the data acquisition. The results are compared to that of DAS based method. 

\textit{In-vivo Carotid Artery}: Finally, the proposed beamforming technique is tested on an \textit{in-vivo} carotid dataset captured from a healthy volunteer by following the principles outlined in the Helsinki Declaration of 1975, as revised in 2000. The instantaneous quiver plots and the velocity profiles along with the velocity estimates over time are compared to that of DAS based methods. 
For all the experimental studies, the TAC estimator is used with NLHR beamforming and DAS beamforming. The parameters used for the \textit{in-vitro} flow phantom experiments and the \textit{in-vivo} studies are shown in Table \ref{tbl1}.
	\subsection{Performance Metrics}
The bias and standard deviation in velocity magnitude ($V_{Bias\%}$ and $V_{SD\%}$) for the simulation and experimental studies, if not reported differently, are determined as:
\begin{equation}\label{eqn10}
    V_{Bias\%}\ =\frac{V_M\ {-\ V}_T}{V_P}\times100\ \ \ ;\ V_{SD\%}=\frac{Std(V_M)}{V_P}\times100	
\end{equation}
where $V_M,{\ V}_T$ and ${\ V}_P$ are the measured, the true (or theoretical), and the peak velocity values, respectively; $Std(\bullet)$ denotes the standard deviation function.
The accuracy of angle estimation is determined using the bias ($A_{Bias}$) and standard deviation ($A_{SD}$) measures as given by \eqref{eqn11}:
\begin{equation}\label{eqn11}
    A_{Bias}\ =A_M\ {-\ A}_T\ \ ;\ A_{SD}=Std(A_M)
\end{equation}
where $A_M$ and ${\ A}_T$ are the measured and the true (or theoretical) values, respectively. Neither spatial and temporal averaging nor any bias compensation is applied to the velocity estimates. It should be noted that the DAS beamforming is performed on the temporally and spatially resampled dataset to ensure a fair comparison.

\begin{figure*}[pos=t]
	\centering
	  \includegraphics[width=\textwidth,height=\textwidth,keepaspectratio]{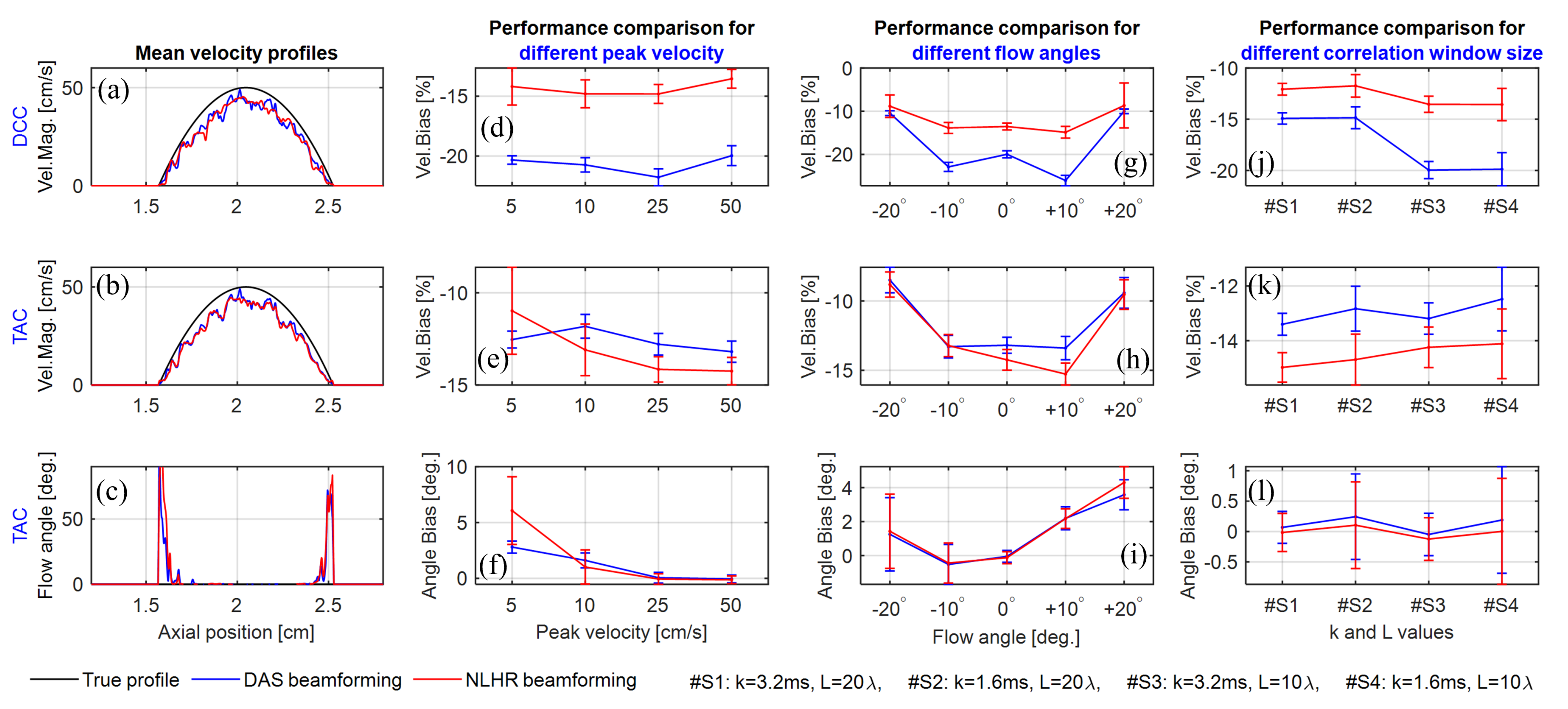}
	\caption{Mean estimated velocity profiles with DAS and NLHR beamforming for DCC estimator (a), TAC estimator (b, c). Quantitative comparison of Bias and Standard Deviation (Std. Dev.) in velocity magnitude and flow angle for test case 1 with DCC (d), and TAC (e, f) estimators. Comparison plots for test case 2 with DCC (g), and TAC (h, i) estimators. Comparison plots for test case 3 with DCC (j), and TAC (k, l) estimators. The error bar is centered at median bias and the length of the error bar corresponds to one standard deviation.}
	\label{fig2}
\end{figure*}


\section{Results}\label{results}
This section discusses the results obtained from the detailed simulation study and the experimental investigations. Detailed video results are attached as supplementary files and are available online [\href{https://www.youtube.com/playlist?list=PLiuuVhVNWBZSYikqhd20FsVr8NTKRlZ4F}{Online Link}].

\subsection{Simulation Results}
The estimated velocity for a transverse parabolic profile of peak velocity $50\ cm/s$ with the proposed NLHR beamforming for DCC estimator compared with that of DAS beamforming is shown in Fig. \ref{fig2}(a) and the comparison plots for velocity magnitude and angle with TAC estimator are shown in Fig. \ref{fig2}(b) and (c) respectively. The quantitative performance comparison of NLHR beamforming based velocity estimations with that of DAS based methods in terms of bias and standard deviation in the magnitude and angle estimates for test cases 1, 2, and 3 are shown in the plots in Fig. \ref{fig2} columns 2, 3, and 4 respectively. For the plots shown in Fig. \ref{fig2} (d)-(l), the inner 90\% region of the vessel radius is considered. In general, an underestimation in the velocity magnitude is observed with respect to the true profile for both the beamformers and estimators. The bias and standard deviation in the velocity magnitude and angle estimates with NLHR beamforming are observed to be similar to that of DAS beamforming for TAC based estimates. However, a reduction in the bias is observed for the velocity magnitude estimates with NLHR beamforming for DCC based estimates whereas the standard deviation is found to be similar, if not better.

\subsection{Phantom Results}
\begin{figure*}
	\centering
	  \includegraphics[width=\textwidth,height=\textwidth,keepaspectratio]{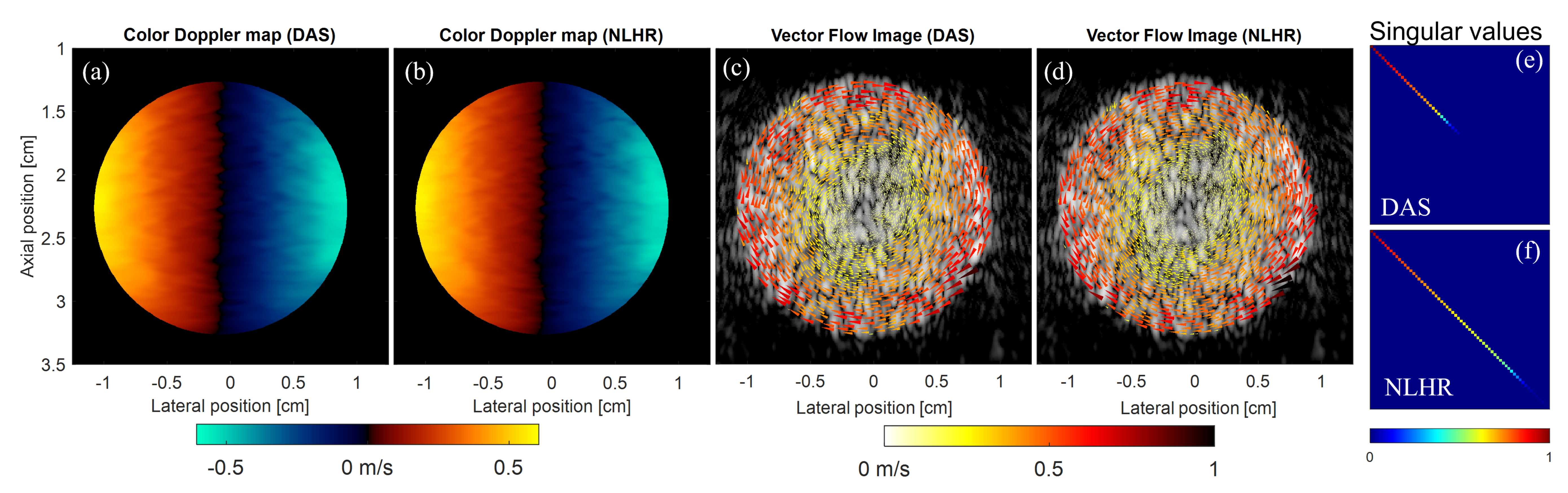}
	\caption{Results for rotating disk dataset. (a), (b) Color Doppler map obtained with DAS and NLHR beamforming respectively. (c), (d) Vector flow images obtained with DAS and NLHR beamforming respectively. (e), (f) Normalized singular values of the beamformed matrix for DAS and NLHR respectively. The B-mode images in (c) and (d) are reconstructed with DAS beamforming. The difference between DAS and NLHR beamforming for rotating disk can be clearly seen from the \textit{disk\_1.gif} and \textit{disk\_2.gif} (supplementary material)}.
	\label{fig3}
\end{figure*}

\textit{In-vitro Rotating Disk:} Clutter filter was not required for this dataset \cite{Posada2016StaggeredDoppler,Madiena2018ColorSampling}. The comparison of the color Doppler plots and vector flow images for DAS and NLHR beamforming is shown in Fig. \ref{fig3}(a)-(b) and (c)-(d), respectively. The major observations from this experiment are: 1) the velocity estimates with NLHR beamforming matches with that of DAS beamforming 2) the color Doppler map of NLHR beamforming is slightly sharper than that of DAS and provide better velocity estimates at the disk edges as evident from the \textit{disk\_1.gif} (supplementary material). The improved spatial localization of the beamformed signals due to the coherence enhancement and the improvement in contrast resolution due to the suppression of incoherent echoes with NLHR beamforming could be seen in the \textit{disk\_2.gif} (supplementary material). The experiment, being an evaluation of different axial and lateral velocities spanning all the flow directions (i.e., $0 - 2\pi$ radians), confirms the feasibility of nonlinear beamforming in flow imaging. \\
\begin{figure*}
	\centering
	  \includegraphics[width=\textwidth,height=\textwidth,keepaspectratio]{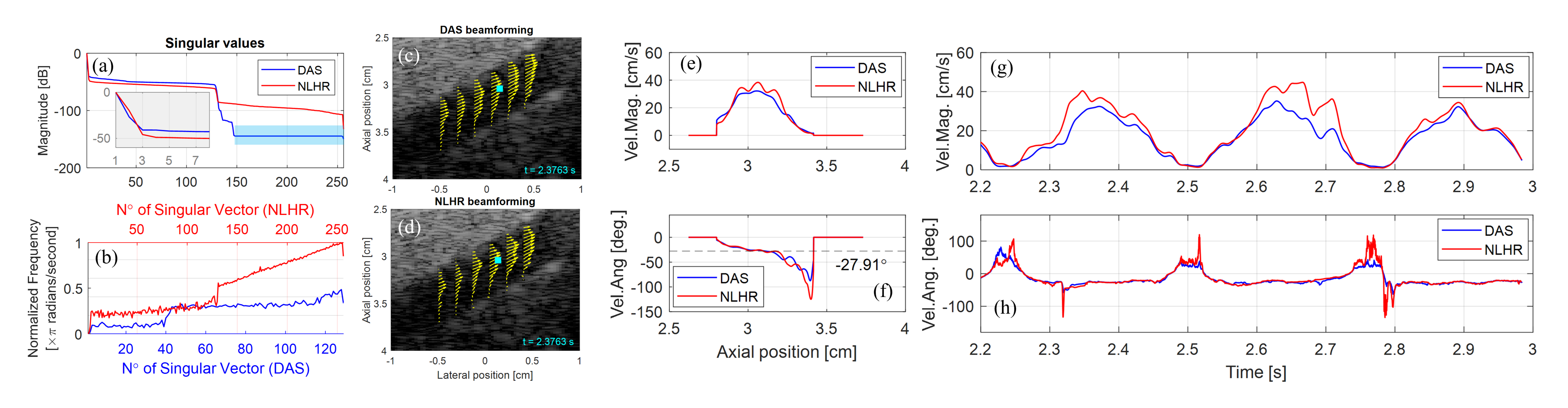}
	\caption{(a) Magnitude of the singular values of beamformed data matrix expressed in dB for the left aperture (at $\alpha=9^\circ$) in DAS and NLHR beamforming (b) Frequency of temporal singular vectors corresponding to the singular values in (a). For a fair comparison, only 128 significant singular vectors are shown for DAS in (b). Vector flow images obtained with DAS beamforming (c) and NLHR beamforming (d) for phantom studies: testcase 1. Velocity magnitude (e) and angle (f) profiles for the axial line passing through the point shown in (c) and (d). The median of estimated angle ($-27.91^\circ$) with NLHR beamforming is shown in dotted line in (f). Velocity magnitude (g) and angle (h) profiles over time at the point shown in (c) and (d) for a typical pulsatile flow. The video result for this test case is available as supplementary material (\textit{phantom\_pulsatileFlow.mp4}).}
	\label{fig4}
\end{figure*}
\begin{figure*}
	\centering
	  \includegraphics[width=\textwidth,height=\textwidth,keepaspectratio]{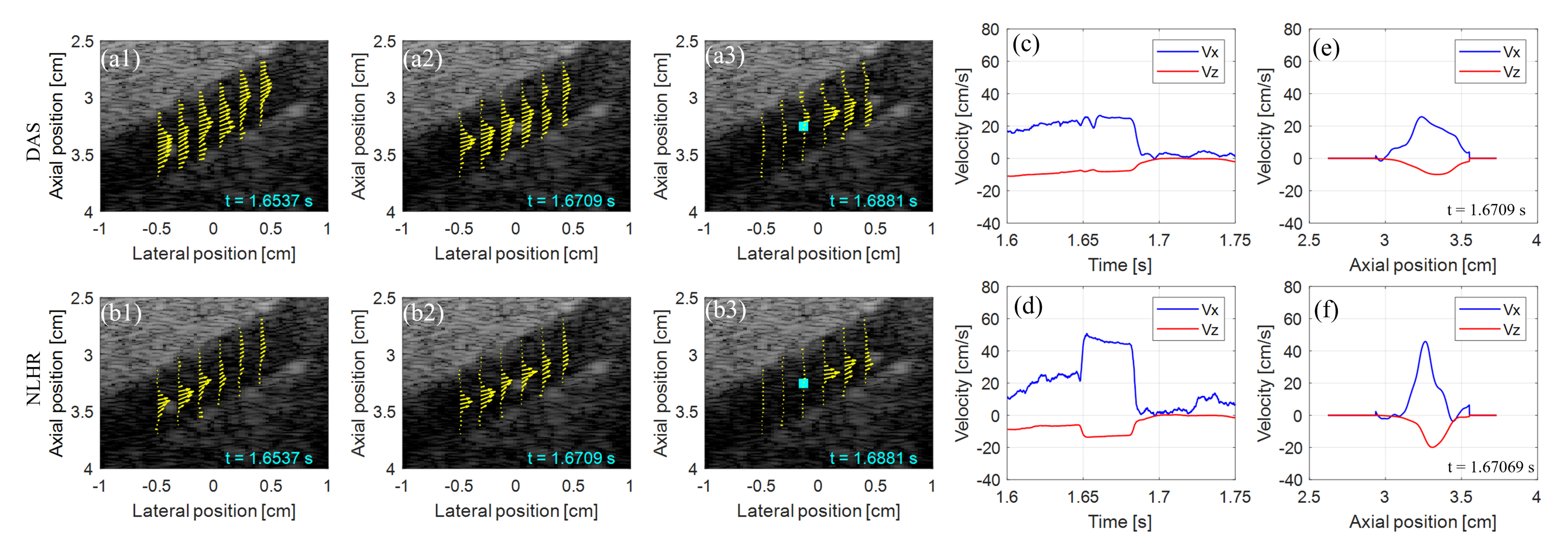}
	\caption{Quiver plots captured during the motion of an air bubble for DAS (a1)-(a3) and NLHR beamforming (b1)-(b3). The time instant at which the frames are captured is shown in cyan at bottom right corner of all the images. The B-mode images in all plots are reconstructed DAS beamforming. (c)-(d) Velocity time trace during the passage of the air bubble obtained with DAS and NLHR beamforming respectively. (e)-(f) Velocity profiles for the axial line passing through the point shown in (a3) and (b3) obtained with DAS and NLHR beamforming respectively. The video result for this test case is available as supplementary material (\textit{phantom\_airBubbleTracking.mp4}).}
	\label{fig5}
\end{figure*}
\begin{figure*}
	\centering
	  \includegraphics[width=\textwidth,height=\textwidth,keepaspectratio]{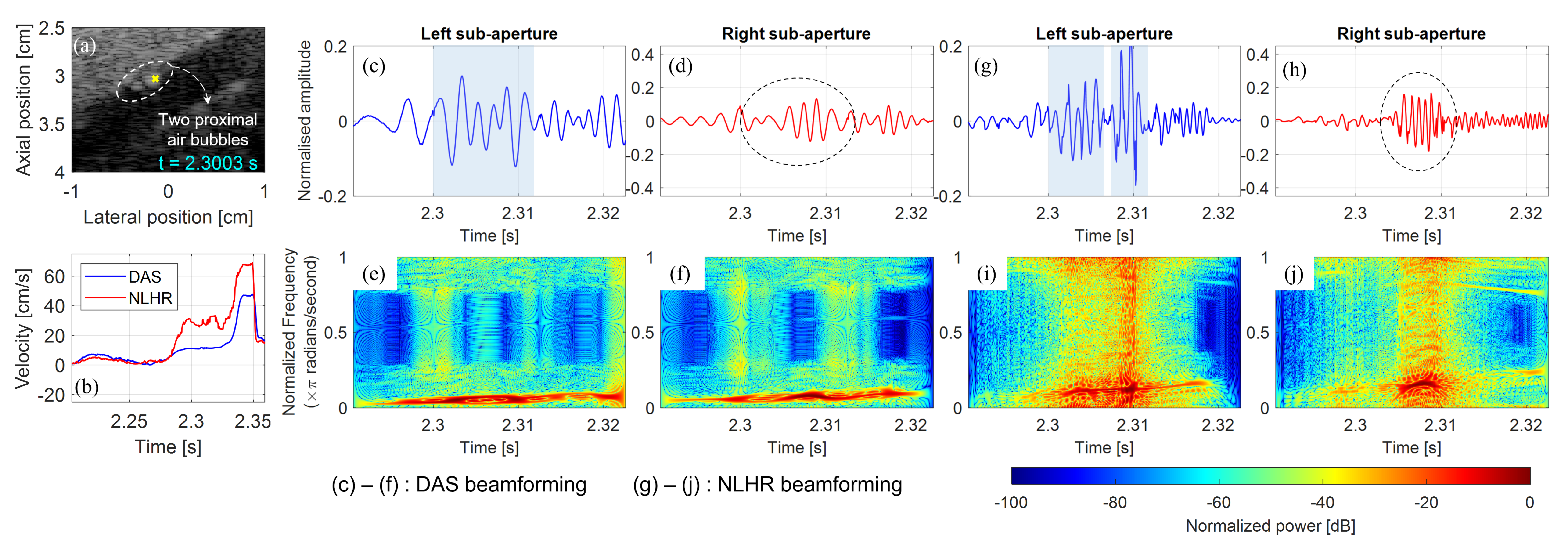}
	\caption{(a) B-mode image showing the two proximal air bubbles in the flow phantom. (b) Comparison of velocity time trace during the passage of the air bubbles with DAS and NLHR beamforming. (c), (d) DAS beamformed slow-time signals for left and right sub-aperture respectively. (e), (f) The time-frequency plots of the DAS beamformed signals in (c) and (d) respectively. (g), (h) DAS beamformed slow-time signals for left and right sub-aperture respectively. (i), (j) The time-frequency plots of the NLHR beamformed signals in (g) and (h) respectively. The plots in (b)-(j) correspond to the pixel indicated in yellow in (a).}
	\label{fig6}
\end{figure*}
\begin{figure}
	\centering
		\includegraphics[width=0.45\textwidth,height=0.45\textwidth,keepaspectratio]{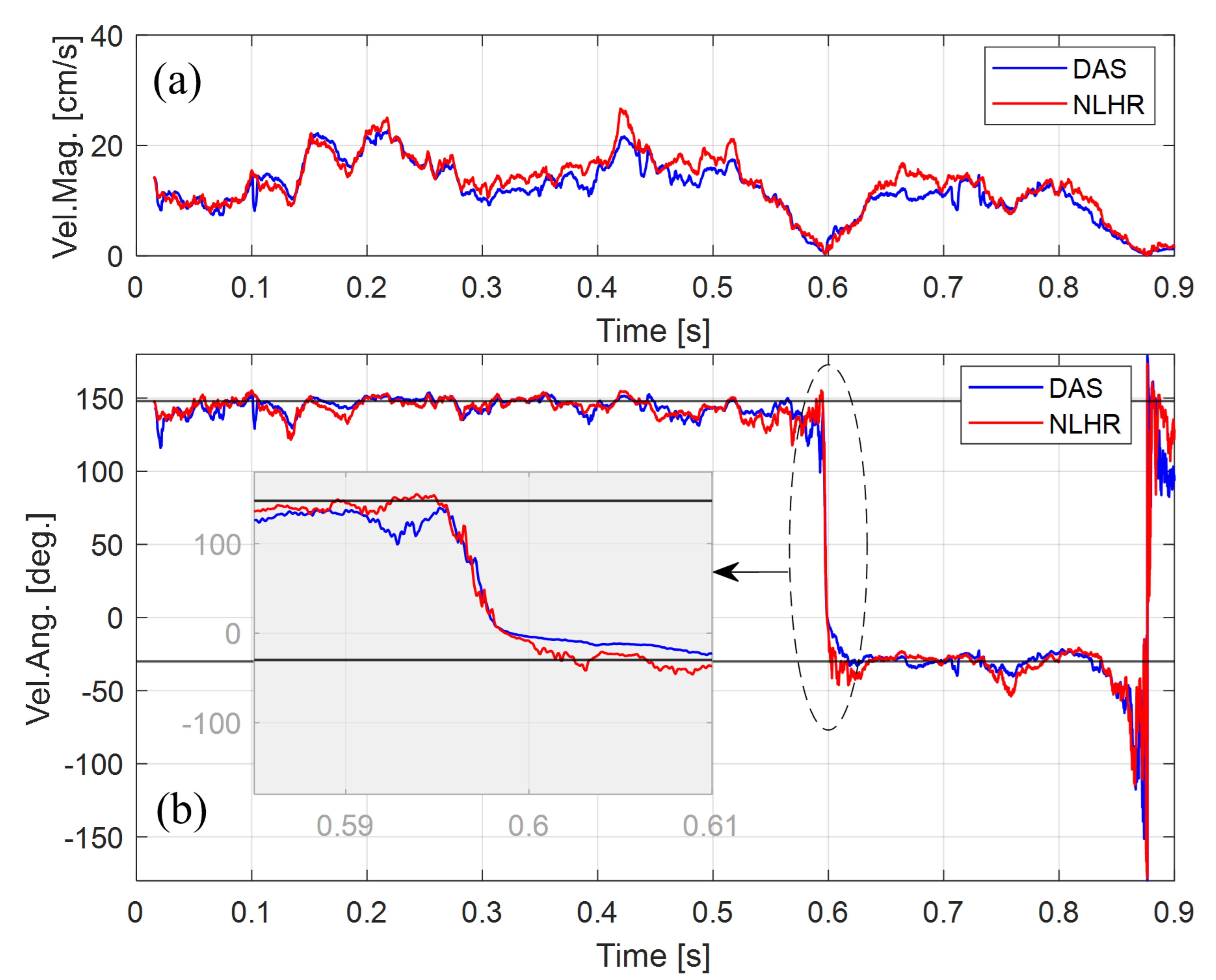}
	\caption{Velocity magnitude (a) and angle (b) profiles over time obtained with DAS and NLHR beamforming for phantom studies: testcase 3. An enlarged view of the angle estimates during the flow direction reversal is shown with in (b). The horizontal black lines indicate the possible flow angles as per the vessel orientation in the phantom. The video result for this test case is available as supplementary material (\textit{phantom\_flowDirectionReversal.mp4}).}
	\label{fig7}
\end{figure}
\begin{figure*}
	\centering
	  \includegraphics[width=\textwidth,height=\textwidth,keepaspectratio]{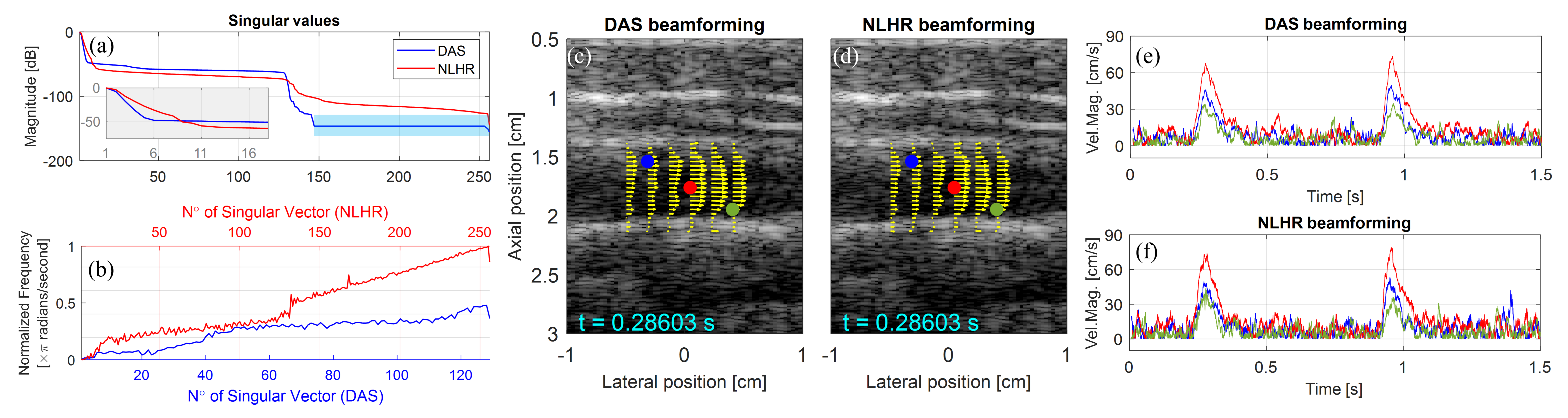}
	\caption{(a) Magnitude of the singular values of beamformed data matrix expressed in dB for the left aperture (at $\alpha=9^\circ$) in DAS and NLHR beamforming (b) Frequency of temporal singular vectors corresponding to the singular values in (a). For a fair comparison, only 128 significant singular vectors is shown for DAS in (b). Quiver plot obtained with DAS beamforming (c) and proposed NLHR beamforming (d) overlaid on the DAS beamformed B-mode image. Mean estimated velocity profiles with DAS based TAC with a correlation window of 16 ms (e) and NLHR based TAC with a correlation window of 16 ms (f) for the points shown in (c) and (d) respectively. The line colors in (e) and (f) corresponds to the dot colors in (c) and (d).}
	\label{fig8}
\end{figure*}
\textit{In-vitro Flow Phantom Experiments:} Unlike simulations and the rotating disk dataset, clutter filtering was required for the flow phantom experiments. The top 3 and top 5 singular value signal components were removed for DAS based estimation and NLHR based estimation respectively to suppress the clutter. The choice of these thresholds is in accordance with the power of the singular values observed for the beamformed signals as shown in Fig. \ref{fig4}(a). A correlation interval of $32\ ms$ (corresponds to 256 frames of acquisition and 512 frames after resampling) is used for velocity estimation with DAS and NLHR based methods. Fig. \ref{fig4}(c)-(h) shows a comparison between DAS and NLHR beamforming for a typical pulsatile flow dataset in which the velocity estimates of DAS and the proposed nonlinear beamforming are observed to be very similar. It is interesting to note that the parabolic nature of the flow is more evident in the NLHR when compared to the slightly flat-topped result in DAS (referring to Fig. \ref{fig4}(e)).

Fig. \ref{fig5}(a1)-(a3) and (b1)-(b3) shows the set of frames (captured at three instances during the flow of an air bubble) for DAS and NLHR beamforming, respectively. A spatially well resolved velocity tracking of the air bubble is observed for NLHR when compared to DAS beamforming. The velocity transients due to the air bubble is captured more concisely by the proposed NLHR beamforming than DAS beamforming as evident from the velocity-time plots in Fig. \ref{fig5}(c) and (d). This is further supported by the axial velocity plot that depicts the velocity variations in the presence of an air bubble as shown in Fig. \ref{fig5}(e) and (f). Thus Fig. \ref{fig5} shows the improved spatial resolution (in terms of localized spatial velocity) and velocity sensitivity (as evident from the localized time variation of velocity). 

To investigate further, we consider a scenario in which two proximal air bubbles are moving in the vessel as shown in Fig. \ref{fig6}(a). From the velocity-time trace during the flow of the air bubbles shown in Fig. \ref{fig6}(b), it is interesting to notice that the NLHR based estimates distinguish between the two air bubbles more distinctly. The beamformed slow time signals of the left and right sub-apertures for the correlation window (of 512 samples) during the flow of air bubbles are shown in Fig. \ref{fig6}(c), (d) for DAS, and in Fig. \ref{fig6}(g), (h) for NLHR beamforming. A clearer separation is observed in the NLHR beamformed left sub-aperture signals as highlighted in Fig. \ref{fig6}(g) when compared to that of DAS in Fig. \ref{fig6}(c). In addition to this, the NLHR beamformed right sub-aperture signals corresponding to the air bubble are observed to be much more confined in time than that of DAS as indicated in Fig. \ref{fig6}(d) and (h). The time-frequency plots of the beamformed slow-time signals in Fig. \ref{fig6}(c), (d), (g) and (h) are shown in Fig. \ref{fig6}(e), (f), (i) and (j). These highly localized time-frequency plots were obtained with the reassigned pseudo Wigner-Ville distribution \cite{Auger1995ImprovingMethod}. The velocity transients, induced by the flow of air bubbles, which are characterized by the excitation of all the frequencies in the spectrum are evident for the NLHR beamforming as shown in Fig. \ref{fig6}(i) and (j) than that of DAS in Fig. \ref{fig6}(e) and (f). 

Fig. \ref{fig7}(a) and (b) summarizes the result for the test case in which a sudden reversal of the flow direction is imposed. The velocity magnitude estimates were observed to be similar for DAS and NLHR beamforming. However, the angle estimates with the proposed NLHR beamforming appear to be  converging with the possible flow angles according to the vessel orientation in the phantom as shown in Fig. \ref{fig7}(b). The video results for all the test cases are available as supplementary files which represent the inferences quite comprehensively. 

\subsection{Preliminary \textit{In-vivo} Validation}
For the \textit{in-vivo} dataset, the top 6 and top 11 singular value signal components were removed for DAS based estimation and NLHR based estimation respectively. The choice of these thresholds is again in accordance with the power of the singular values observed for the beamformed signals as shown in Fig. \ref{fig8}(a). We also observe that the singular vectors corresponding to those singular values are associated with the tissue and wall motions that are characterized by low frequency as shown in Fig. \ref{fig8}(b). The quiver plots obtained with DAS and NLHR beamforming during peak systole are shown in Fig. \ref{fig8}(c) and (d) respectively. The velocity-time trace for three different locations (one in the middle of the vessel and the other two closer to the vessel wall as indicated in Fig. \ref{fig8}(c) and (d)) in the carotid artery is shown in Fig. \ref{fig8}(e) and (f). The video result with more details of the \textit{in-vivo} study is available as  supplementary material (\textit{invivo\_carotid.mp4}).

\section{Discussion}\label{discussion}
As discussed in Section \ref{Introduction}, the limited sensitivity of the conventional flow imaging techniques were addressed either with coherent compounding techniques or an improved clutter filter design. In this paper, we attempt to introduce a novel multiply and sum based nonlinear beamformer in flow imaging that could provide better spatiotemporal sensitivity towards the flow transients. The multiply and sum operation in F-DMAS beamforming is proven to provide enhanced coherence in the beamformed signals with significantly better contrast and lateral resolution in B-mode imaging \cite{Matrone2015TheImaging}. Owing to these advantages of multiply and sum operation, the proposed approach for flow imaging has shown  better performance in terms of spatial resolution and velocity sensitivity as compared to state-of-the-art DAS based methods. The output of each step involved in the proposed nonlinear beamformer for an \textit{in-vitro}
rotating disk dataset is included as supplementary material. 

\subsection{Better Spatial Resolution and Velocity Contrast}\label{discussionA}
The improvement in the spatial resolution is observed with the spatial distribution of the velocity vectors in Fig. \ref{fig5}, where the air bubble is tracked along with the typical blood flow velocity. This improved velocity contrast is due to the improved contrast resolution which is a characteristic property of DMAS beamformer. In line with the characteristic of F-DMAS in \cite{Matrone2015TheImaging}, the following characteristics can be observed for the proposed non-linear beamformer as well in flow imaging: 
\begin{enumerate}
    \item The widening of the “synthetic” receive aperture due to the correlation function in (\ref{eqn4}) and (\ref{eqn5}) that has $\ 2N_c-1$ coefficients as in the case of the F-DMAS beamformer in \cite{Matrone2015TheImaging} 
    \item The doubling of the center frequency of the beamformed signal  provides twice the number of frequency components than in a DAS beamformed signal (as evident from the singular value plots in Fig. \ref{fig3}(e), (f), Fig. \ref{fig4}(a), (b) and Fig. \ref{fig8}(a), (b)) that could be considered as the improvement in frequency resolution or equivalently the velocity resolution.
\end{enumerate}

Conversely, the signal components in DAS beamformed signal for the highlighted portion in Fig. \ref{fig4}(a) and Fig. \ref{fig8}(a) are arguably not relevant for velocity estimation for two reasons: 1) the frequency of all those components are observed to be the same and 2) the energy of those components are observed to be the same and very minimal. Moreover, this signifies that the closer frequencies (Doppler shifts) are better resolved with NLHR beamforming than DAS and a greater number of spatiotemporal components will be available for the SVD clutter filtering resulting in more efficient clutter suppression when compared to that of DAS. The improvement in the velocity and the spatial resolution may not be significant in typical parabolic flow evaluations as we have observed with the simulations (Fig. \ref{fig2}) and experiments (Fig. \ref{fig3} and Fig. \ref{fig4}). However, the improved velocity contrast and spatial resolution are highly appreciated in the presence of flow transients as demonstrated \textit{in-vitro} in Fig. \ref{fig5} and Fig. \ref{fig6}.

\subsection{Enhanced Sensitivity Towards Flow Transients}
Flow velocity transients (ideally, impulsive changes in the velocity) are characterized by the excitation of almost all the frequencies in the spectrum of the beamformed signals. The introduction of the air bubbles to the Doppler fluid would effectively induce such velocity transients and it is interesting to note that the proposed beamformer was sensitive enough to capture the impulsive transients as evident from the time-frequency plots in Fig. \ref{fig6}. Moreover, the comparison of beamformed signals (Fig. \ref{fig6}(c) and (g)) and its time-frequency plot suggests that the transients created by each of the air bubbles are better resolved (temporally) with the proposed nonlinear beamforming. This is due to the coherence enhancement and the harmonic generation with themultiply and sum operation. Also, better flow angle estimates with the NLHR beamforming observed during the flow direction reversal (Fig. \ref{fig7}) further signifies the sensitivity of the proposed approach towards the flow transients (both in terms of the magnitude and direction of the flow).  

\subsection{Applications} 
The proposed technique with improved spatial resolution and velocity contrast would be of great relevance in microbubble localization and tracking, remote drug delivery, neovascularization, and microvasculature imaging applications. Furthermore, the enhanced sensitivity of the proposed approach could probably help in identifying new biomarkers or pathological features in medical diagnostics. 

\subsection{Limitations and Future Scope}
The advantages of the proposed approach are at the cost of increased computational complexity when compared to DAS beamforming. So the proposed approach may not be suitable for portable imaging systems in its current form and would require further optimization. To address the computational complexity of the proposed approach, neural network accelerators which have found success in B-mode image reconstruction will be employed to speed up the proposed workflow as an immediate future work. Further, the proposed approach must be investigated with detailed \textit{in-vivo} animal studies. 

Since the major objective of this work is to improve the spatiotemporal sensitivity in high frame rate flow imaging, the proposed technique is demonstrated and validated only for non-steered plane transmit schemes and  other transmit schemes like focused, synthetic aperture, and diverging wave transmit were not considered during the formulations of the proposed approach. Hence the proposed nonlinear approach must be re-visited for the transmit schemes other than plane waves. In this regard, further developments of this work also include the analysis of this approach for synthetic aperture and diverging wave vector flow imaging and power Doppler. 

With reference to the improvement in the velocity contrast offered by the proposed nonlinear beamformer, as discussed in Section \ref{discussionA}, the proposed approach can provide a better separation between closer velocities. This would have a significant impact on the evaluation of slow flows and the velocity estimates towards the periphery of blood vessels because such low-velocity magnitudes overlap with that of the tissue motion. In this regard, the scope of this work could be further expanded to a systematic study on the impact of the proposed nonlinear beamforming in the design of clutter filter. 

\section{Conclusion}\label{conclusion}
Inspired by the advantages of F-DMAS beamforming in B-mode imaging, this article investigates the performance of a nonlinear beamformer in high frame rate ultrasound flow imaging. Towards this, a novel multiply and sum based nonlinear beamformer is introduced for high frame rate flow imaging and is demonstrated using DCC and TAC based velocity estimators. Further, the proposed approach has been investigated in detail for velocity sensitivity with \textit{in-vitro} datasets including a rotating disk, air bubble tracking, and flow direction reversal. The proposed approach has performed similar, if not better, for parabolic flow simulations and typical pulsatile flows \textit{in-vitro} and in preliminary \textit{in-vivo} studies. However, owing to the enhanced coherence and the harmonic generation by the multiply and sum operation in the beamformer, an enhanced velocity contrast, spatial resolution, and sensitivity towards the flow transients were observed \textit{in-vitro}. The results suggest that the method has the potential to greatly impact the studies on microbubble tracking, cavitation, and other applications in which velocity sensitivity and resolution are of prime interest. However, the improved spatiotemporal sensitivity offered by the proposed beamformer is at the cost of increased computational complexity and hence the proposed approach in its current form may not be suited for portable flow imaging systems that are constrained by power and computational resources. Moreover, in its current form, we have not considered any transmit schemes other than plane waves, and hence more studies are required to investigate the feasibility of the proposed approach for other transmit schemes.

\section*{Supplementary Materials}
The detailed video results for the various test cases are attached as supplementary files and are available online at \url{https://shorturl.at/fzCF9}. The MATLAB code of the proposed algorithm is available at \url{https://github.com/madhavanunni/Nonlinear_Beamforming.git}. 

\section*{Declaration of Competing Interests} 
The authors declare that they have no known competing financial interests or personal relationships that could have appeared to influence the work reported in this paper.

\section*{Acknowledgement}
The authors would like to acknowledge the funding from the Department of Science and Technology - Science and Engineering Research Board (DST-SERB (ECR/2018/ 001746)) and the Ministry of Human Resource Development (MHRD), India. The authors are grateful to Ms. Gayathri M and Mr. Pisharody Harikrishnan G for providing adequate help in data acquisition
and the authors also acknowledge the high-performance computing facility provided by Indian Institute of Technology Palakkad, Kerala, India. The authors would also like to gratefully acknowledge the comments from the anonymous reviewers that contributed significantly to the improvement of this manuscript.

\printcredits
\bibliographystyle{unsrtnat}

\bibliography{references.bib}

\end{document}